\documentclass[twoside,11pt]{article}
\usepackage{asp2004}
\usepackage{psfig}
\usepackage{graphics}

\begin{document}

\title{A proposed origin for chondrule-forming shocks in the solar
nebula}

\author{Andrew F. Nelson}
\affil{Dept of Physics and Astronomy,
       202 Nicholson Hall,
       Louisiana State University, 
       Baton Rouge LA, 70803, USA}

\vskip 1mm

\author{Maximilian Ruffert}
\affil{School of Mathematics, 
      University of Edinburgh, 
      Edinburgh EH9 3JZ, UK}

\begin{abstract}
We propose that the nebular shocks currently favored as a model to
form chondrules and other annealed silicates in the solar nebula
originate in the dynamical activity present in the envelope of forming
Jovian planets. In contrast to the classic `core accretion model', our
3D hydrodynamic simulations show that this envelope is not a 1D
hydrostatic structure but is instead vigorously active and contains
densities and temperatures that appear similar in magnitude and
spatial extent to those thought to be responsible for the production
of chondrules. 
\end{abstract}

\vspace{-5mm}

\section{Introduction}

According to the most likely theory for Jovian planet formation,
Jupiter formed in a three stage process, lasting about 6~Myr. In the
classic model \citep{PP4_WGL}, a 5-15$M_\oplus$ rock/ice core forms at
a distance of $\sim$5~AU from the central star \citep{Boss95} over a
period of about 1/2 million years. It then begins to grow a
hydrostatic, gaseous envelope which slowly cools and contracts while
continuing to accrete both gas and solids until the total
core$+$envelope mass reach about 30 $M_\oplus$, after another $\sim6$
Myr have passed. In the final stage, the envelope begins to collapse
and a period of relatively rapid gas accretion follows, ending with
the planet in its final morphology. As stated, the timescale presents
a considerable problem for the model because circumstellar disks are
observed \citep{HLL01} to survive for only $\sim4$~Myr, though a large
dispersion in age remains and a few survive until much later. More
recent models \citep{IWI03,AMB04,HBL04} cut the timescale to
$\sim1$~Myr by invoking additional physical processes such as
migration or opacity modifications in the material in the forming
envelope. 

A critical assumption in all versions of the core accretion model is
that the gaseous envelope is hydrostatic. We present a study designed
to investigate whether this assumption is in fact valid, and to
investigate the existence and character of the activity in the flow if
it is not. Our motivation for this study is to begin an exploration of
the possibility that the core accretion timescale may be further
shortened by the dynamical activity without the costs associated with
the other recent models. After finding that the flow is indeed quite
active, we propose that one consequence of the shocks resulting from
the activity is the production of chondrules and other annealed
silicates in the solar nebula.

\section{Initial Conditions and Physical Model}

We simulate the evolution of the gas flow in a 3 dimensional (3D)
Cartesian cutout region of a circumstellar disk in the neighborhood of
an embedded Jovian planet core. We derive the initial conditions in a
two stage process. First, we define a set of global conditions for the
disk as a whole, then we extract a small volume for which we simulate 
the evolution.

The global conditions are similar to those described in \citet{JovI}
and  assume that the disk orbits a 1$M_\odot$ star modeled as a point
mass. The disk extends from 0.5 to 20 AU and is described by surface
density and temperature power laws, each proportional to $r^{-1}$. We
assume that at an orbital radius of $a_{\rm pl}=5.2$~AU the surface
density and temperature are $\Sigma_{\rm pl}=500$~gm~cm$^{-2}$ and
$T_{\rm pl}=200$~K. We define the orbital velocities such that
centrifugal accelerations are exactly balanced by the combined
pressure and gravitational accelerations from the star and the disk.
Radial and vertical velocities are set to zero. With these dimensions,
the total implied disk mass of our underlying global model is
$M_D\approx.035M_\odot$. At the core's orbit radius, the implied
isothermal scale height is $H=c_s/\Omega\approx0.40$~AU, where $c_s$
and $\Omega$ are the local sound speed and rotation frequency
respectively. The disk is stable against the growth of gravitational
instabilities as quantified by the well known Toomre $Q$ parameter,
which takes a minimum value of $Q\approx5$ near the outer edge of the
disk. In the region near the core's orbit radius, its value is $Q>15$.

To simplify the local initial condition, we neglect the $z$ component
of stellar and global disk self gravity, but include the full
contribution of the core's gravity and local disk self gravity,
defined as the component of disk gravity originating from matter
inside our computational volume. This simplification allows us to
neglect the disk's vertical structure. Since we expect that the most
interesting part of the flow will be confined to the volume in and
near the core's Hill sphere, and both the grid dimensions and the disk
scale height are significantly larger, neglecting the vertical
stratification will have only limited impact on our results.

The origin of our coordinate grid is centered on a 10$M_\oplus$ core,
orbiting the star at $a_{\rm pl}=5.2$~AU. We use a modified `shearing
box' approximation to translate the cylindrical equations of motion
into the rotating Cartesian frame. Our modification includes
non-linear terms neglected in the standard form of
\citet{GLB-shearsheet}, allowing a closer correspondence between the
global and local conditions. Our modification allows the shear in the
$x$ direction, corresponding to the radial coordinate, first, to
include a non-zero offset of the corotation radius from the core's
position and, second, does not need to vary linearly with $x$,
as occurs when pressure contributes to the disk's rotation curve.

We extract the local initial condition from the global condition by
mapping the radial and azimuth coordinates of the two dimensional
global initial condition directly onto the $x$ and $y$ coordinates of
the local Cartesian grid, centered on the core, using the mapping: 
$x= r - a_{\rm pl}$ and $y = r\phi$. Quantities in the $z$ direction
are obtained by duplicating the midplane quantities at each altitude.
The $x$ and $z$ velocities are defined to be zero. We obtain the $y$
velocity by subtracting off the orbital motion of the core from the
azimuth velocity at each radius and mapping the remainder into our
Cartesian grid at the appropriate $x$ position. 

Although we avoid complications associated with modeling the disk's
vertical structure because we neglect the $z$ component of stellar and
disk gravity, we still require a correspondence between the globally
defined disk surface density and the locally defined volume density
used in the actual calculations. To make the connection, we use the
conversion $\rho=\Sigma/H$, where $\rho$ and $\Sigma$ refer to the
volume and surface densities respectively, and the isothermal scale
height $H=c_s/\Omega$. This conversion introduces a small physical
inconsistency, since of course our physical model omits the physics
responsible for producing vertical structure in the first place. The
inconsistency means that the volume density will contain a small
systematic error in its value, however since the exact value of the
volume density in the Jovian planet environment is not well known, we
believe this inconsistency will not be important for the results of
our simulations. 

We use an ideal gas equation of state with $\gamma=1.42$ and include
heating due to compression and shocks, but no radiative heating or
cooling. This value of $\gamma$ is chosen to be representative of
values found in the background, solar composition circumstellar disk,
for which temperatures imply that the rotational modes of hydrogen
will be active. We expect the gas to be optically thick in the region
of interest, so that thermal energy will remain with the fluid rather
than being radiated away. The core is modeled as a point mass, onto
which no gas may accrete. The conditions at the boundaries are fixed
to the values of the global initial condition, resulting in a steady
flow into and out of the grid that mimics the near-Keplerian flow of
the underlying circumstellar disk.

On a global scale, the disk will respond only weakly to the influence
of a 10$M_\oplus$ core and will never form a deep gap
\citep{DHK02,JovI}, so a time varying boundary condition is not
required. A more serious concern is whether the flow within the
simulation volume becomes sufficiently perturbed away from that inside
the boundaries, to cause an unphysical back reaction to develop. We
have monitored the flow for signs of such effects and have found that
for the simulation presented here, perturbations have become well
enough mixed with the background flow so that quantities near the
boundaries are not altered substantially from their initial values. We
believe effects from numerical perturbations of this sort will have
minimal impact on the results. We caution that we have not found the
same statement to be true throughout our parameter space, e.g., at
very low background temperatures.

We use a hydrodynamic code \citep{Ruf92} based on the PPM algorithm
\citep{ColWood84}, which has been adapted to use a set of nested grids
to evolve the flow at very high spatial resolution. Both smooth and
shocked flows are modeled with high accuracy because PPM solves a
Riemann problem at each zone interface to produce the fluxes at each
timestep. Shocks and shock heating are therefore included as an
integral part of the method. No additional artificial viscous heating
is required or included. Each successive grid is overlaid on top of
the central portion of its parent grid, but with one half of its
linear dimensions. Each grid in the nest contains an identical number
of zones, so that the effective linear spatial resolution is doubled
in the overlay region. In the model presented here, we use a nest of
six grids. The simulation volume extends 4 Hill radii ($R_H=a_{\rm
pl}(M_{\rm pl}/3M_\odot)^{1/3}$, corresponding to about $1.1H$) in
each direction, defining a volume of (0.897 AU)$^3$. Regions both
above and below the disk midplane are included in the simulation
volume. The finest grid covers $\pm1/8R_H$ in each direction with a
spacing of $\sim6.5\times 10^9$~cm per zone, corresponding to about
1.3 times the diameter of Neptune.

\vspace{-5mm}

\section{Results of our simulations} 

We have performed a large set of simulations covering a range of both
initial conditions and physical models and a paper describing each of
these studies is in preparation. For our purposes, it is sufficient to
summarize the results by examining one model in detail, whose initial
conditions were described in the last section, and which was run for a
total of 100~yr of simulation time. We consider it to be the most
realistic model of those we studied in the sense that it includes the
most complete inventory of physical processes. 

\begin{figure} 
\psfig{file=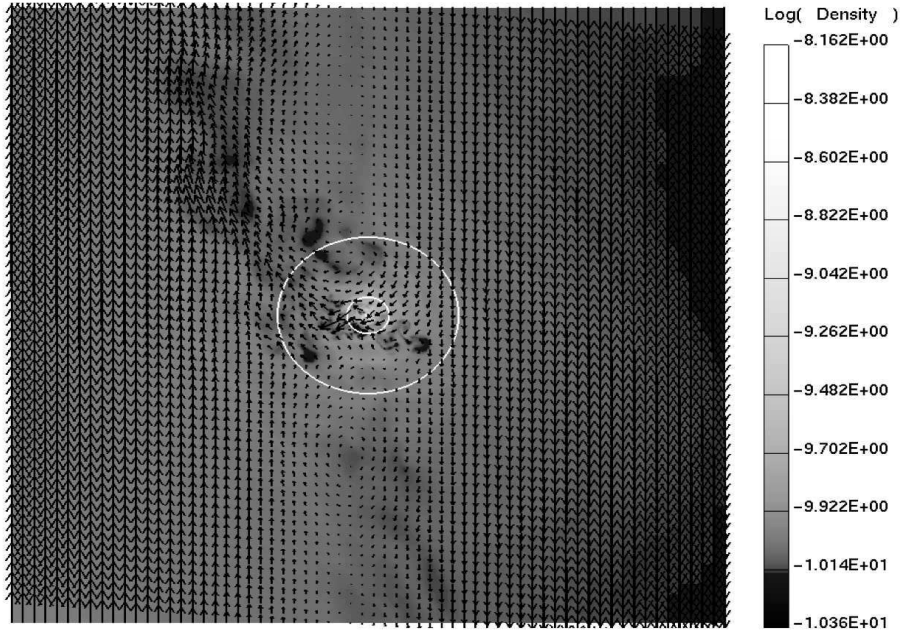,height=95mm,width=120mm,angle=-90}
\psfig{file=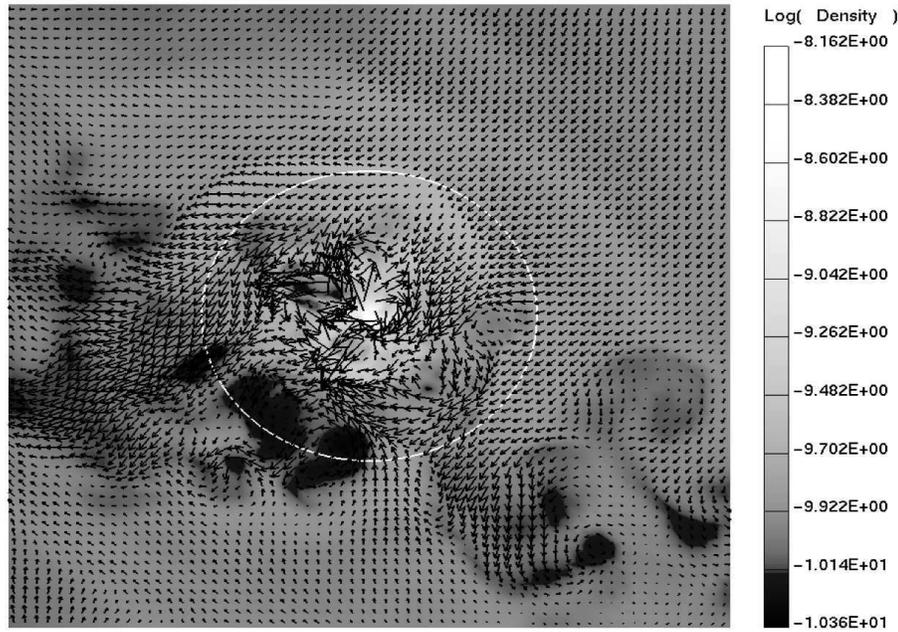,height=95mm,width=120mm,angle=-90,rheight=91mm}
\caption{\label{fig:cutout-mid-dens}
The volume density in a 2D slice taken through the disk midplane for
the full simulation volume (top), and a blowup of the region within
$\pm1/2R_H$ of the core (bottom). Velocity vectors are shown projected
onto the plane on the coarsest grid in the top panel, and on the
fourth nested grid in the bottom panel. The white circles define the
radius of the accretion sphere $R_A=GM_{\rm pl}/c^2_s$ (small circle)
and the Hill radius (large circle). The grey scale is logarithmic and
extends from $\sim10^{-10}$ to $\sim10^{-8}$ gm~cm$^{-3}$. } 
\end{figure}

\begin{figure} 
\begin{center}
\psfig{file=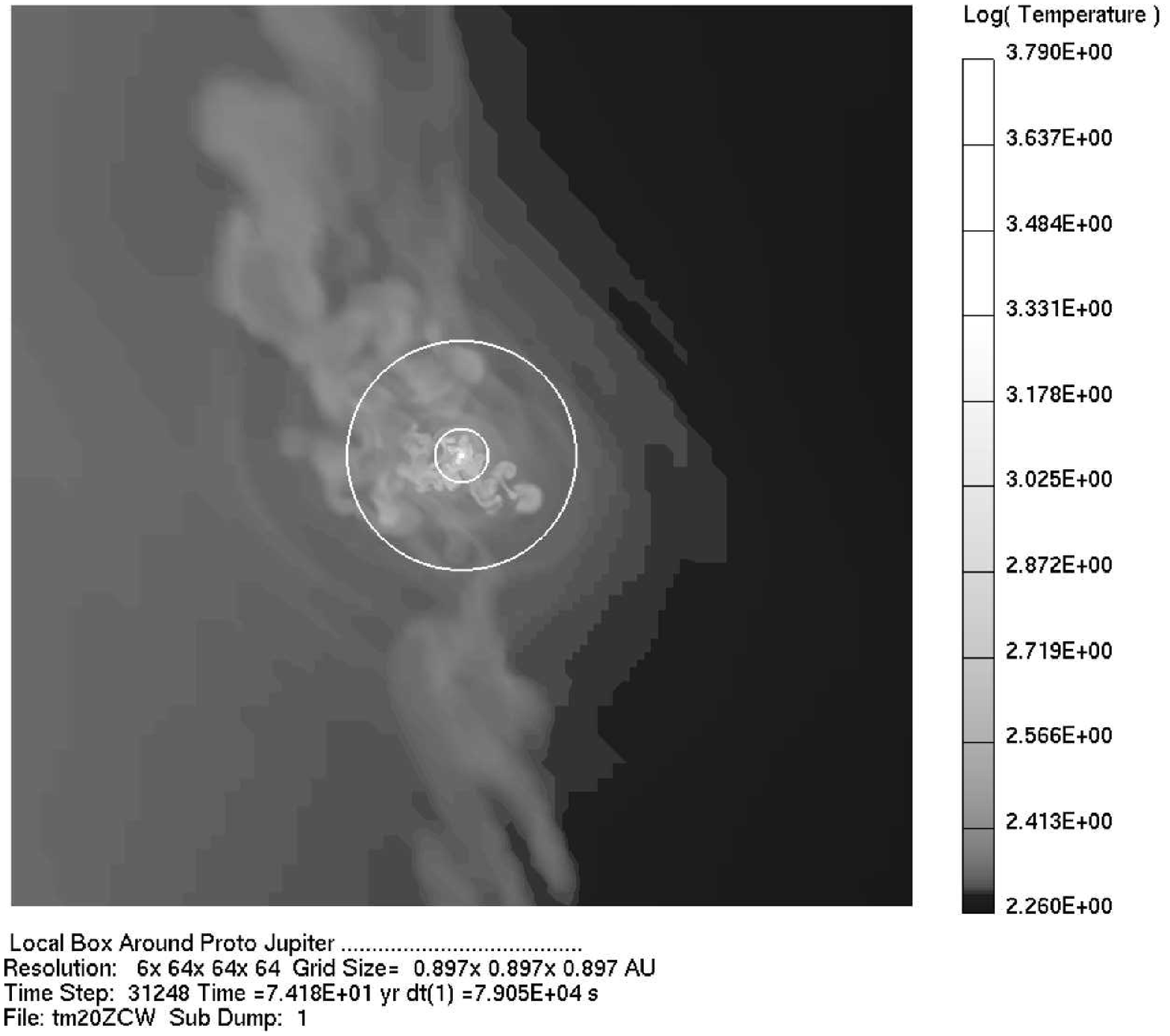,height=100mm,width=120mm,angle=-90}
\psfig{file=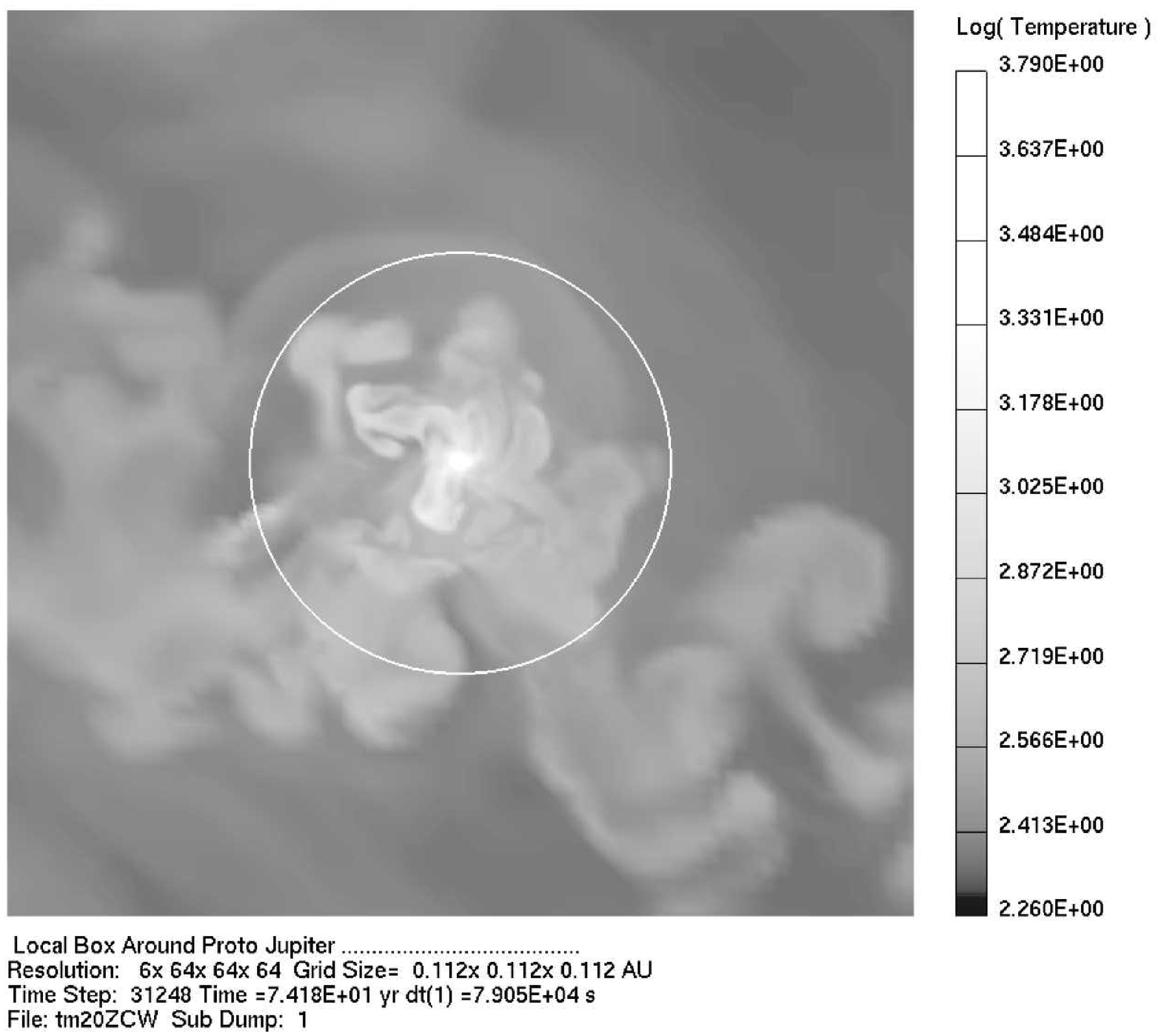,height=100mm,width=120mm,angle=-90,rheight=90mm}
\end{center}
\caption{\label{fig:cutout-mid-temp}
As in Figure \ref{fig:cutout-mid-dens}, but showing temperature. The
color scales are logarithmic and extend from 180 to 6500 K.
Temperatures as high as 3-5000~K are common very close to the core,
with temperatures decreasing rapidly to the background $\sim$ 200~K
value at increasing distance.}
\end{figure}

In Figures \ref{fig:cutout-mid-dens} and \ref{fig:cutout-mid-temp}, we
show 2D slices of the gas density and temperature, taken through the
disk midplane at a time 74~yr after the beginning of the simulation.
In both Figures, the structures are highly inhomogeneous and become
progressively more so closer to the core. Densities both above and
below that of the background flow develop due to shocks that produce
hot `bubbles', which then expand into the background flow. One such
bubble is particularly visible in the plots of the temperature
distribution, emerging to the lower right. Such structures are common
over the entire the duration of the simulation and emerge in all
directions, depending on details of the flow at each time. Activity
persists for the entire simulation, and for as long as we have
simulated the evolution without significant decay or growth. Lower
resolution models that were run for much longer ($\sim1600$~yr) also
display continuing activity. However, since we neglect cooling, we
cannot expect the flow to become much less active over time.

In conflict with the expectation from orbital mechanics that the flow
of material approaching the Hill volume will turn around on a
`horseshoe' orbit, matter approaching the outer portion of the Hill
volume is relatively unaffected, often passing completely through its
outer extent with only a small deflection. In contrast with this quiet
flow further away, material is very strongly perturbed on the scale of
the accretion radius, where large amplitude space and time varying
activity develops. This too conflicts with the orbital mechanics
picture, in which matter inside the Hill volume simply orbits the
core. Material can enter the Hill volume from the background flow and
shocked, high entropy material can escape and rejoin the background
flow. Changes in the flow pattern occur on timescales ranging from
hours to years, with a typical encounter time inside the accretion
radius of less than a month.

\vspace{-5mm}

\section{The new scenario for chondrule formation}\label{sec:chond}

As readers of this proceedings volume will be aware, the theory of
chondrule formation suffers from no lack of data, but rather from
insufficient understanding of what physical processes are present in
the solar nebula, where they are present and whether they produce
conditions appropriate for chondrule formation. Briefly summarized
from \citet{PP4_Jones}, we note for our purposes that chondrules
underwent both a very rapid heating event to $\sim$2000~K, followed
quickly by a rapid cooling event of magnitude 50-1000~K/hr. Among a
veritable zoo of models purporting to produce such conditions, passage
of solid material through nebular shocks is currently favored as among
the most likely (see e.g., Desch et~al., in these proceedings).
Among its drawbacks are, first, that shocks that have the right
density, temperature and velocity characteristics are hard to form
and, second, that it is difficult to arrange for these shocks to exist
for a long enough time to produce enough bodies to match the current
observations. 

The parameter space for which chondrule production may occur in shocks
\citep[][Table 5]{DC02} is bounded by preshock densities within a
factor of a few of $10^{-9}$~gm~cm$^{-3}$, temperatures near 300~K and
shock speeds near 6--7~km~s$^{-1}$, and that enhancements in the
particle density were important for formation models. Concurrent work
of \citet{CH02,iida01} come to similar conclusions. Figures
\ref{fig:cutout-mid-dens} and \ref{fig:cutout-mid-temp} show that
appropriate background conditions exist in our simulations, and we
propose that dynamical activity in the Jovian envelope could provide a
source for both shocks and reversible compressive heating that remedy
the shortcomings noted above. 

\begin{figure}
\psfig{file=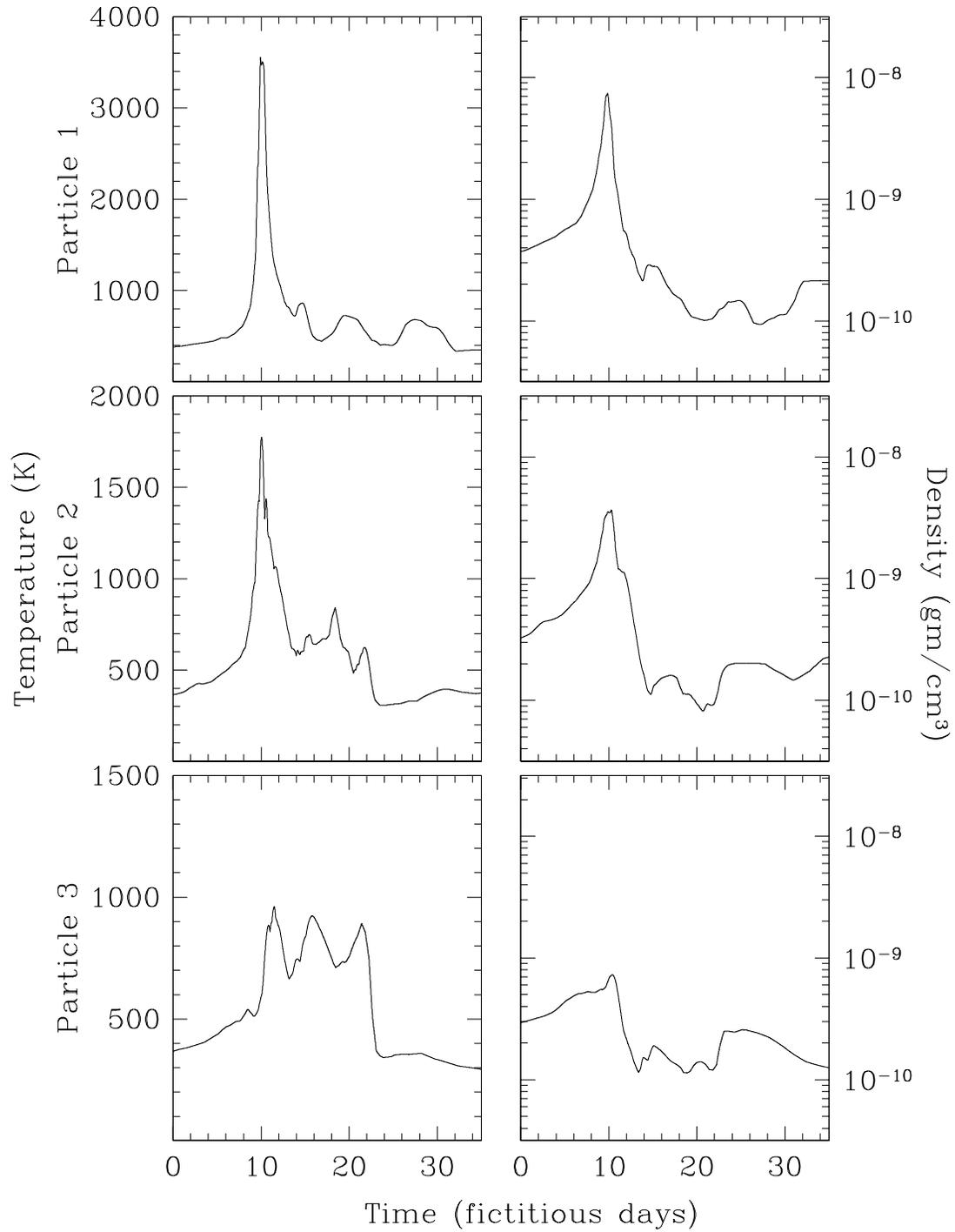,height=8.0in,width=6.0in,rheight=7.7in}
\caption{\label{fig:thermo-traj}
The temperature (left panels) and density (right panels) encountered
by three test particles as they are advected through a snapshot of the
simulation volume, each as functions of a fictitious time coordinate.
The zero point for time has been arbitrarily shifted in each case so
that the peak temperature occurs at $\sim10$~days.}
\end{figure}

Although the basic temperature and density conditions can be seen
in the Figures, ascertaining whether short duration heating and
cooling events are also present requires additional analysis. We
considered and discarded the option of including test particles
that could be passively advected with the local flow, because of
the significant added computational and storage cost they would
require. Instead, we rely on the similar but not identical solution
of `advecting' test particles through a snapshot of the conditions
at a specific time.  

To that end, we have performed a streamline analysis of the
trajectories of an ensemble of test particles injected into the
simulation restart dump for the same time as that shown for Figures
\ref{fig:cutout-mid-dens} and \ref{fig:cutout-mid-temp}. Particles are
placed at a set of locations at the edge of the grid and allowed to
advect through the simulation volume using the local fluid velocity,
linearly interpolated between the grid zones adjacent to the particle
at each time. It is given a timestep based on that used to advance the
gas at that location, so that it advances forwards through a
fictitious time coordinate, passing through the volume at the rate of
matter in the local flow. Similar linear interpolations of density and
internal energy are used to derive the temperature from the equation
of state. 

In Figure \ref{fig:thermo-traj}, we show temperatures and densities
for three test particles for which a passage through the environment
of the core has occurred. The particles shown were chosen to illustrate
the range of peak temperatures and densities that may be
encountered by slightly different trajectories through the envelope.
Each encountered a short duration heating and cooling event as they
passed through the dynamically active region close to the core, but
the magnitudes and duration varied in each case. In the top example
(test particle 1), the peak temperature rose to well over 3000~K and
the density to nearly $10^{-8}$~gm~cm$^{-3}$ as the particle's trajectory
passed through the innermost regions of the envelope. The conditions
encountered by the particle 2 were much more moderate, with peak
temperature and density values of $\sim1800$~K and
$4\times10^{-9}$~gm~cm$^{-3}$ respectively. Particle 3's trajectory took
it only through the outer portion of the envelope, so that it
encountered only much lower temperatures and densities, although in
this case for a much longer period of time than either of the other
two cases shown. All three particles encountered several days of
cooler processing near 500-800~K, and a visual scan of many other
similar events shows that such additional annealing is not uncommon. 

The temperature peaks for test particles 1 and 2 offer widths of $\la
1$ day, with both a very rapid rise and fall. Close examination of
their trajectories reveal that the widths reflect essentially the
crossing time for a single grid zone in the simulation. Therefore,
although already quite narrow, we believe that they are actually
overestimates of the true widths that would be obtained from the
models as realized at still higher resolution. The dynamical activity
in the envelope, coupled with the temporally very narrow
temperature/density peaks, especially in cases similar to that of
particle 2, offer evidence that chondrule production could occur
in the environment of Jovian planet formation. 

\vspace{-5mm}

\section{Concluding comments, questions and skeptical remarks} 

The scenario we present offers a number of attractive advantages over
other models for shock formation in the solar nebula. First, it
naturally provides a mechanism for producing very short duration
heating and cooling events with thermodynamic characteristics similar
to those expected to produce chondrules. Unlike models using global
gravitational instabilities in the circumstellar disk, it does not
require a massive disk for the activity to exist, and in particular,
for a massive disk to continue to exist for the long period of time
required to produce chondrules in large quantities. Production in this
scenario will endure for a significant fraction of the formation
timescale for Jovian planets (itself a significant fraction of the
disk lifetime), resulting both in a large yield of objects and
allowing both processing and reprocessing events to occur. Also,
because there were a number of similar proto-Jovian objects in the
solar system, processing will occur in many locations. If correct, our
results mean that the standard core accretion model for Jovian planet
formation will require significant revision, and will imply both
link between the timescales for chondrule formation and planet
formation, and that chondrules represent a physically examinable link
to the processes present during the formation of Jovian planets.

There are still many unanswered questions contained in this scenario,
however. Before any detailed analysis of the conditions will be of
real use for either the theory of chondrule formation or planet
formation, we must perform simulations that include both radiative
transport and a non-ideal equation of state for the gas. Without them,
the densities and temperatures obtained in our simulations will
contain significant deviations compared to the real systems they are
intended to model. Moreover, including them means the dynamical
properties of the system will change, perhaps eliminating the shocks
altogether. Preliminary indications with locally isothermal and
locally isentropic equations of state suggest that at least some
activity will remain, so we remain hopeful. 

We have simulated only a 100~yr segment of the Jovian planet formation
history, during a time when the envelope did not contain much mass. We
cannot be certain that the activity will remain when the envelope
becomes more massive. 

If we find that shocks are produced in more physically inclusive
models, it will be interesting to perform a significantly more
detailed analysis of the conditions in those shocks, including their
velocities relative to the fluid flow. Will such analysis show that
the shocks fit into the required density/temperature/velocity
parameter space? One concern already apparent is that the flow
velocities of material flowing through the shocks (1-2 km~s$^{-1}$, as
estimated from the directly available fluid flow velocities
themselves) are uncomfortably low compared to those quoted by
\cite{DC02} and \cite{iida01}. It seems unlikely that the
velocities will be increased as dramatically as that by any of the
improvements to the models we might make.

Although the results from our streamline analysis are promising, they
are no substitute for an investigation of the trajectories of specific
packets of material through system. A detailed study of this issue
will be important on several levels. First, it is not clear that a
particle's thermodynamic trajectory will be the same when it is
advected through an actual time dependent flow, as opposed to the
fictitious advection through a fixed flow that we have performed. It
will also be important to understand what fraction of material that
approaches the core actually encounters conditions appropriate for
chondrule formation during its passage, in comparison to material that
instead encounters regions that are inappropriate. From a slightly
broader perspective, the same question becomes what fraction of the
total budget of solid material in the solar nebula undergoes such
processing? Secondly, after ejection from the envelope, it will be
important to understand how the processed materials get from where
they form (near 5 AU) to their final locations, in meteorites
throughout the inner solar system.

Finally, in our discussion we have focused solely on the conditions
required for the production of chondrules. On the other hand, they are
not the only meteoritic material that has undergone heating and
cooling events. \citet{HD02} discuss one such class of material,
composed of annealed silicates found in comets, for which the required
temperatures and densities are much lower. Will our scenario be able
to produce material of this sort as well? 

In future work, we plan to implement successively more advanced models
to simulate the Jovian planet formation process. One important aspect
of this project will be to address questions important for the
formation of chondrules and other annealed silicates in much more
detail than our current models allow.

\acknowledgements{
AFN gratefully acknowledges support from NSF grant NSF/AST-0407070,
and from his UKAFF Fellowship. The computations reported here were
performed using the UK Astrophysical Fluids Facility (UKAFF). AFN
gratefully acknowledges conversations with A. Boley and R. Durisen
during the conference that led to the streamline discussions
surrounding Figure \ref{fig:thermo-traj}. }


\vspace{-5mm}

\begin{thebibliography}{}

\bibitem[Alibert {et al.\ }(2004)]{AMB04} Alibert, Y, Mordasina, C. \&
Benz, W. 2004, 417, L25

\bibitem[Boss(1995)]{Boss95} Boss, A.P. 1995, Science, 267, 360

\bibitem[Ciesla \& Hood(2002)]{CH02} Ciesla, F. J. \& Hood, L. L 2002,
Icarus, 158, 281

\bibitem[Colella \& Woodward(1984)]{ColWood84} Colella P. \&
Woodward, P. R. 1984, J. Comp. Phys., 54, 174

\bibitem[D'Angelo {et al.\ }(2002)]{DHK02} D'Angelo, G., Henning,
T. \& Kley, W. 2002, A\&A, 385, 647

\bibitem[Desch \& Connolly(2002)]{DC02} Desch, S. J. \& Connolly, H. C.
Jr. 2002, Meteoritics \& Planetary Science, 37, 187

\bibitem[Goldreich \& Lynden-Bell(1965)]{GLB-shearsheet} Goldreich,
P. \& Lynden-Bell, D. 1965, MNRAS, 130, 125

\bibitem[Harker \& Desch(2002)]{HD02} Harker, D. E. \& Desch, S. J.
2002, ApJ 565, 109

\bibitem[Haisch {et al.\ }(2001)]{HLL01} Haisch, K. E., Lada, E. A. \&
Lada, C. J. 2001, ApJL, 553, 153

\bibitem[Hubickyj {et al.\ }(2004)]{HBL04} Hubickyj, O., 
Bodenheimer, P. \& Lissauer, J. J. 2004, RMxAC, 22, 83

\bibitem[Iida {et al.\ }(2001)]{iida01} Iida, A, Nakamoto, T. \&
Susa, H. 2001, Icarus, 153, 430

\bibitem[Inaba {et al.\ }(2003)]{IWI03} Inaba, S., Wetherill,
G. W. \& Ikoma, M. 2003, Icarus, 166, 46

\bibitem[Jones {et al.\ }(2000)]{PP4_Jones} Jones, R. H., Lee,
T., Connolly, H. C. Jr., Love, S. G. \& Shang, H. 2000, In Protostars
and Planets IV, pp. 927-962, ed. Mannings, V., Boss, A. P. \&
Russell, S. S., University of Arizona Press: Tucson

\bibitem[Nelson \& Benz(2003)]{JovI} Nelson, A. F. \& Benz, W. 2003,
ApJ, 589, 556

\bibitem[Ruffert(1992)]{Ruf92} Ruffert, M. 1992, A \& A, 265, 82

\bibitem[Wuchterl {et al.\ }(2000)]{PP4_WGL} Wuchterl, G.,
Guillot, T. \& Lissauer, J. J. 2000, In Protostars and Planets IV, 
pp. 1081-1110, ed. Mannings, V., Boss, A. P. \& Russell, S. S., 
University of Arizona Press: Tucson

\end{thebibliography}
\end{document}